\documentstyle[twoside,fleqn,espcrc2]{article}
{\newcommand{\lsim}{\mbox{\raisebox{-.6ex}{~$\stackrel{<}{\sim}$~}}}
{\newcommand{\gsim}{\mbox{\raisebox{-.6ex}{~$\stackrel{>}{\sim}$~}}}
{\newcommand{\gtwid}{\mathrel{\raise.3ex\hbox{$>$\kern-.75em\lower1ex
\hbox{$\sim$}}}}
{\newcommand{\ltwid}{\mathrel{\raise.3ex\hbox{$<$\kern-.75em\lower1ex
\hbox{$\sim$}}}}

\newcommand{\AmS}{{\protect\the\textfont2
  A\kern-.1667em\lower.5ex\hbox{M}\kern-.125emS}}

\hyphenation{nucleo-syn-the-sis}
\hyphenation{re-commend-ed}

\title{Primordial Baryon Asymmetry and Sphalerons}

\author{Kimmo Kainulainen\address{Department of High Energy Physics (SEFL),
        P.O. Box 9 (Siltavuorenpenger 20 C),\\
        FIN-00140, University of Helsinki, Finland}
        \thanks{Address after 1.1.1995, Theory Division, CERN, CH-1211,
                Geneva 23, Switzerland}}

\begin{document}

\begin {abstract}
I show that a cosmological baryon asymmetry generated at the GUT scale
is in general safe against washout due to sphalerons and generic $B$- or
$L$-violating effects.  This result is mainly due to the (almost) conserved
number of right-handed electrons at high temperatures $T \gsim {\cal
O}(10) $ TeV, but also the mass corrections, in particular the thermal
masses of leptons act as the protector of the primordial baryon
asymmetry.
\end{abstract}

\maketitle
\section{Introduction}

It was realized a long time ago that the evident excess of
baryons over antibaryons in the universe could be generated during the
early stages of its evolution by generic out of equilibrium $B$- and
$CP$-violating interactions \cite{sakharov}. First quantitative
implementations of these
ideas came in the context of grand unified theories (GUTs), where the out
of equilibrium conditions, $B$-violation and a large enough $CP$-violation
were  rather easily realized \cite{fot}. Later other scenarios of
baryon asymmetry  generation that work in the very early universe have
been proposed \cite{fy1}. It took surprisingly long time before it
was realized that  all the conditions required for baryon number
generation are qualitatively satisfied in the Standard Model during the
electroweak phase transition\cite{rev}.  Whether the electroweak
baryogenesis can be made to work quantitatively remains still an open and
undoubtedly one of the most challenging questions  in the modern physics.

Standard model baryogenesis became a possibility when it was understood
that the anomalous baryon number violating sphaleron interactions are
unsupressed  at high temperatures \cite{sph}. This very same phenomenon
however, appeared to imply the death of any primordial (as opposed to the
electroweak) baryogenesis mechanisms. Roughly speaking this is because
sphaleron interactions destroy any net excess in $B+L$-number, whereas
e.g.\ the simplest GUTs predict
$B-L=0$; combining these one immediately gets $B=L=0$. A way out would seem
to be offered by more complicated models, like SO(10)-GUT
which can generate a nonzero $B-L$. Such models, however, necessarily
contain a number of nonrenormalizable $B$- and/or $L$-violating
interactions. The danger of such interactions to primordial baryon
asymmetry was first noticed by Fukugita and Yanagida (FY) \cite{fy} who
observed that the lepton number  must be at least a fairly good
approximate symmetry. Their reasoning was that  if the lepton violating
processes were significant, then effectively $L$ would be driven to zero
and since sphalerons cause $B+L \rightarrow 0$, one is again reduced to
having zero baryon asymmetry.

It should be noted that the presence of $B$ and/or $L$-violating
interactions does not automatically imply the vanishing of baryon number;
it is possible to adjust all the parameters of the theory in such a way
that these interactions are too weak to affect the asymmetries
significantly. In essence the GUT baryogenesis could then be made to work
if certain set  of ``consistency constraints'' between the parameters of
the theory were satisfied. In the light of the subsequent developments
these constraints appeared to become rather strong however, so as to
make the primordial baryogenesis much more unappealing. Indeed, since the
observation made by FY a great deal of effort has been spent to
strenghten their result
\cite{ht,nb,sonia} and to generalize it to other baryon and/or lepton
number violating operators \cite{sonia,cdeo12}. The strongest bounds were
obtained by requiring that the $B$- and/or $L$-violating interactions
had to be out of equilibrium ever since the temperature at which
the sphalerons first came into equilibrium at $T_m \sim 10^{12}$ GeV,
rather than the much lower electroweak phase transition temperature used
by FY.

The situation changed again after a somewhat surprising result was pointed
out in ref.\ \cite{cko23}: most of the efforts to strenghten the  original
FY-type constraint are invalidated by a rather mundane feature  of the
Standard Model, namely the smallness of the Yukawa coupling of the
right-handed electron.\footnote{In the supersymmetric case
it was realized earlier \cite{iq}
that above a certain scale associated with supersymmetry breaking, $T_s
\sim 10^8$ GeV, the presence of new anomalies would cause the baryon
number to be encoded in supersymmetric particles, saving it from erasure
until temperatures below $T_s$.}  The key observation is that any
$e_R$-asymmetry remains untouched until around
$T_*\simeq {\cal O}(10)$ TeV, when the small Yukawa interactions with
left-handed electrons and Higgs bosons finally become fast enough to
convert the
$e_R$'s into $e_L$'s. Because sphalerons interact only with the
left-handed particles, they can only directly deplete the latter.
Therefore, as long as any additional lepton and/or baryon violating
interactions have gone out of thermal equilibrium before the right-handed
electrons come {\it into} equilibrium, the initial $e_R$ asymmetry is
protected from being washed out. When the temperature eventually falls
below $\sim {\cal O}(10)$ TeV, sphalerons will be able to convert a sizeable
fraction of the initial $e_R$ asymmetry into the baryon excess that exists
today.

The new conservation law of $e_R$-number above $T_*$ has important
consequences for the evolution of the primordial baryon number.  Firstly,
it tremendously weakens the consistency constraints on the parameters of
the theories with initial $B-L \neq 0$. This will be demonstrated below
for the particular lepton number violating operator considered by FY.
Secondly, with theories having initially $B-L=0$, the old vice turns into
a new virtue; any $B$  and/or $L$-violating operator in equilibrium at $T_*
\lsim T \lsim 10^{12}$  GeV would, with rapid sphaleron transitions and
$e_R$-conservation,  help to {\it generate} a baryon asymmetry. This can
occur because the rapid $B$ and/or $L$-violation imposes new equilibrium
between the chemical potentials of the interacting species which
necessarily corresponds to $(B-L)_{eq} \neq 0$, because $e_R$-species
does not partake into these processes and yet carries a charge quantum
number. When the fermion number violation ceases the asymmetries evolve
in the standard way but with a new initial condition
$(B-L)_i \rightarrow (B-L)_{eq}\neq 0$.

On the basis of the above discussion one can identify two possible
loopholes \footnote{In addition to the obvious possibility of having
yet new exotic interactions that wold bring $e_R$-species into
chemical equilibrium with the left-chiral world.} that could lead to the
destruction of primordial baryon asymmetry. The first is that the $B$-
and/or $L$-violation persists below $T_*$, but this would require
adjusting the parameters of  the theory to get {\it into} the trouble, not
vice versa!  The second is more important: if one has a theory with
$(B-L)_i = 0$ and {\it no} additional fermion number violation, then the
anomalous situation would persist only until $T_*$, after which
the equilibrium with $B=L=0$ would be re-established \cite{cdeo3,cko23}.
However, even then a fraction of the primordial baryon number is
restored due to the finite mass effects. Depending on the order of the
phase transition this would be due to vacuum mass effects \cite{krsdr} or
due to the thermal mass corrections \cite{dko,gs}.

\section{Examples of ${\bf L}$- and ${\bf B}$-violation}
\subsection{Neutrino see-saw mass}

The lepton number violating $\Delta L = 2$, $D=5$ operator
first considered by FY can be written as:
\begin{equation}
{\cal O}_5 = \frac{1}{v^2}\sum_{ij}m_{ij}(\bar L_iH)(H^TL^c_j).
\label{effective}
\end{equation} Here $v=246$ GeV is the usual higgs vev and $m$ is the
see-saw mass matrix of light neutrinos. The rate of lepton number violation
induced by (\ref{effective}) scales with the temperature cubed,
$\Gamma_{\Delta L} \sim T^3$. Hence it is more effective at higher
temperatures and drops out of equilibrium at low enough $T$. An accurate
expression for the rate $\Gamma_{\Delta e_L}$ by which $e_L$-type lepton
number asymmetry is destroyed was computed in \cite{cko23}:
\begin{equation}
\Gamma_{\Delta e_L} = {9\over \pi^5}{T^3\over v^4} \mu^2,
\label{gammaL}
\end{equation}
where $\mu^2 \equiv \frac {5}{3}|m_{ee}|^2+|m_{e\mu}|^2+|m_{e\tau}|^2$.
Comparing this rate to that of the Hubble expansion ($\cong
17 T^2/M_P$)  yields a freezeout temperature of
\begin{equation}
T_f =174({\rm keV}^2/\mu^2){\rm TeV}.
\label{freezeout}
\end{equation}
FY's original constraint $\mu \lsim $ 50 KeV essentially results from
equating $T_f$ with the weak scale although their computation of the rate
(\ref{gammaL}) was less accurate. The improved constraints found in
\cite{ht,nb,sonia} were obtained by equating $T_f$ with the sphaleron
equilibration temperature $T_m \sim 10^{12}$ GeV.

The rate of $e_R-e_L$ interactions on the other hand is determined by
the Higgs decays and inverse decays as well as scattering processes such
as $t_R\bar t_L\to e_R\bar e_L$ and $e_L H\to e_R W$ etc. The rate of
these interactions scale as $\Gamma_{LR} \sim T$ so that they are more
effective at {\it low} $T$ and are out of equilibrium at high $T$.
The total rate of all processes was computed in \cite{cko23} and it was
found to correspond to an equilibration temperature
\begin{equation}
T_* \simeq 1.3f(x) \rm TeV,
\label{Tstar}
\end{equation}
where the function $f(x) \simeq 1.0 + (-1.1+3.0x) + h_t(0.6-0.1x)$
depends on the thermal higgs mass $x\equiv m_H(T)/T$ and the top quark
Yukawa coupling $h_t$; assuming that $m_h=60$ GeV and $m_t=174$ GeV
gives $x\simeq 0.6$ and $T_*\simeq 3$ TeV \cite{cko23}. In any
case $T_*$ is significantly below the sphaleron
equilibration scale $\sim 10^{12}$ GeV and relatively close to the weak
scale. Using $T_* = 3$ TeV one finds
\begin{equation}
\mu \simeq (\theta_{e\mu}^2 m_{\nu_\mu}^2 + \theta_{e\tau}^2
m_{\nu_\tau})^{1/2} \lsim 8 {\rm\ keV},
\label{newbound}
\end{equation}
where the matrix element $m_{ee}$ constrained by the double beta decay
experiments to $m_{ee} < 1$ eV was neglected and the remaining elements
$m_{e\mu}$ and $m_{e\tau}$ were related to the observable mixing angles
and mass eigenstates. One quickly realizes that the bound (\ref{newbound})
must already be satisfied  due to other laboratory and cosmological
constraints \cite{cko23} and thus presents no danger to the primordial
baryon asymmetry.

Suppose now that $L$ violation is occurring above $T_*$ and below $10^{12}$
GeV.  Under these conditions, the $L$-violating and sphaleron reactions
will establish equilibrium between all the interacting species, with the
boundary condition that the $e_R$ asymmetry is conserved. Because $e_R$
carries charge, this constraint carries over to the interacting species
because the universe is charge-neutral. One can easily show that above
$T_*$, $(B-L)_{eq} = -\frac{3}{10}L_{e_R,p}$, where the sub $p$ refers to
the primordial value of the quantity, {\em regardless} of the initial
values of $B$ and $L$. Assuming there was no lepton number violation below
$T_*$, one obtains the final baryon asymmetry according to the standard
analysis \cite{ht}
\begin {equation}
B_f = \frac{28}{79}(B_*-L_*) \simeq -0.11 L_{e_R,p}.
\label{Bfinal}
\end{equation}
Eq.\ (\ref{Bfinal}) holds independent of the initial $B-L$ and in
particular for $(B-L)_i=0$! The strength of $L$-violation needed is quite
modest: all we require is that either an $L$-violating decay or scattering
remain in equilibrium to temperatures below $T_m \sim 10^{12}$ GeV.  This
could be accomplished  by a tau neutrino mass of at least $10^{-2}$ eV.

\subsection{${\bf n-\bar{n}}$-oscillation}
As the second example consider the $\Delta B=2$, $D=9$ operator that would
induce the neutron antineutron oscillations:
\begin{equation}
{\cal O}_9 = \frac{1}{M^5}({\bar u}_R{d_R}^c)({\bar d}_R{d_R}^c)({\bar
u}_R{d_R}^c).
\end{equation} This operator obviously conserves the differences between
the leptonic asymmetries, in particular $C_p \equiv 2L_e - L_\mu - L_\tau$
and imposes a constraint on the chemical potentials $2\mu_{u_R} +
4\mu_{d_R} = 0$. One then readily finds the equilibrium value $(B-L)_{eq} =
\frac{3}{68}(C_p - 9L_{e_R,p})$, valid for $T\gsim T_*$, and the final
baryon asymmetry
\begin{equation}
B_f = \frac{21}{134}(C_p - 9L_{e_R,p}).
\label{nnequ}
\end{equation}
The important thing to notice here is that in
order to generate a nonzero $B_f$ in this example it was not necessary
even to have a primordial asymmetry in the $e_R$-species, all that was
needed was the {\it conservation} of $e_R$-number.

Finally, using $T_* \sim 3$ TeV from (\ref{Tstar}) one finds that the
limit on the heavy  mass scale $M$ suppressing the operator becomes $M
\gtwid 1 \times 10^5$ GeV, which is comparable to the current experimental
bound of $M > 10^5 - 10^6$ GeV.

\section{Finite mass effects}

The zero result obtained for the final baryon asymmetry in the $(B-L)_i=0$
case is an artifact of the massless approximation. Mass effects enter
through the conserved global charge densities that appear
as boundary conditions for the network of equilibrium equations for
chemical potentials. For example, above the electroweak phase transition
one might have
\begin{equation}
Q_{em} = 0,\qquad Q_3 = 0, \qquad B-L = 0,
\label{chdens}
\end{equation}
where $Q_{em}$ is the electric charge density and $Q_3$ is the
isospin density. In the massless approximation these charges are directly
related to the chemical potentials. These relations, however, get
corrections due to finite vacuum masses and due to
thermal interactions. For example
\begin {equation}
Q_{em} = \sum_f \frac{\mu_fT^2}{6}(1-\frac{3x_f^2}{\pi^2}) + ...,
\label{charge}
\end{equation}
where the ellipses refer to the bosonic contribution and e.g.\ above
the phase transition $x_f = m_f(T)/T$ \cite{dko}. It appears natural to
expect that these perturbations should bring about nonzero final
asymmetries roughly of the order of perturbations. This indeed turns out
to be the case with three important refinemets: {\it (i)} corrections in
the quark sector alone are not sufficient because all the quarks have
equal chemical potentials, {\it (ii)} one must have unequal corrections
between lepton families, so that only Yukawa couplings contribute and {\it
(iii)} one must have nonzero differences between the leptonic asymmetries
(these differences are conserved). Then, if the phase transition is 1st
order, the contribution from $T>T_*$ equilibrium values get frozen to the
universe implying
\cite{dko}
\begin{equation}
B_f \simeq  3 \times 10^{-7} (\Delta L_{e\tau} + \Delta L_{\mu \tau}),
\label{basymm}
\end{equation}
where $\Delta L_{ij} \equiv L_i - L_j$. If the electroweak phase
transition is of second order, then the relevant equilibrium is
the one at the broken phase right after the phase transition.
Nevertheless, even then a final baryon asymmetry of the same order
as in (\ref{basymm}) results due from the {\it vacuum} mass effects
\cite{krsdr,dko}.

In conclusion, due to the approximate $e_R$-conservation and the finite
(thermal or vacuum) mass effects the mechanisms of primordial baryon
asymmetry generation remain completely viable in explaining the origin of
the baryon asymmetry in the universe.

\newpage
}}

\end{document}